\pdfoutput=1
\documentclass[reprint,amsmath,amssymb,aps,prb,floatfix,raggedbottom]{revtex4-2}
\usepackage{graphicx}
\usepackage{siunitx}
\usepackage{bm}
\usepackage{xcolor}
\usepackage[colorlinks,allcolors=blue]{hyperref}
\graphicspath{{figs/}}
\newcommand{\ellz}{\langle\ell_z\rangle}
\newcommand{\sz}{\langle s_z\rangle}

\begin{document}

\title{Symmetry-enabled anomalous near-field spin-to-orbital angular momentum conversion}

\author{Tianle Cheng}
\affiliation{School of Power and Mechanical Engineering, Wuhan University, Wuhan 430072, China}
\author{Jinsheng Lu}
\thanks{Contact author: jinslu@whu.edu.cn}
\affiliation{School of Power and Mechanical Engineering, Wuhan University, Wuhan 430072, China}
\affiliation{School of Integrated Circuits, Wuhan University, Wuhan 430072, China}

\begin{abstract}
In the near field of a nanostructure illuminated by circularly polarized light, part of the spin angular momentum of the light is converted into orbital angular momentum, which always has the same sign as the incident spin when the structure has continuous rotational symmetry. Here we show that breaking this symmetry makes the conversion anomalous, reversing the orbital angular momentum of the near field. For silicon nanopillars in water, full-wave calculations show that the near-field orbital angular momentum of a circular pillar follows the incident spin and is enhanced at its resonances, whereas triangular, square, and pentagonal pillars and closely spaced clusters of circular pillars reverse it in distinct spectral bands. The reversal occurs when higher-order components with opposite angular momentum, allowed only by the lower symmetry, dominate over those whose angular momentum has the sign of the spin. These components can be carried by modes of the circular pillar that circularly polarized light cannot excite as long as the continuous rotational symmetry is preserved. These findings show that nanostructure symmetry, not only light polarization, can set the direction of optical orbital angular momentum at subwavelength scales.
\end{abstract}

\maketitle

\section{Introduction}\label{sec:intro}

Light carries spin angular momentum (SAM), associated with circular polarization, and orbital angular momentum (OAM), associated with a phase that winds around an axis \cite{allen1992orbital,yao2011orbital}. Both can be transferred to matter. Circularly polarized light rotates particles about their own axes, and light with a helical phase front drives their orbital motion around the beam axis \cite{he1995direct,yao2011orbital,arzola2019spin,han2018crossover,parker2020optical}. The two forms are not independent, because focusing, scattering, and evanescent fields convert part of the spin into OAM, an effect known as spin-to-orbital angular momentum conversion \cite{zhao2007spin,bliokh2010angular,rodriguezherrera2010optical,bliokh2015spin}.

The conversion is strongest close to nanostructures, where the field varies on subwavelength scales and has large longitudinal components. High-index dielectric nanostructures enhance these fields through electric and magnetic Mie resonances \cite{kuznetsov2016optically,staude2017metamaterial,kruk2017functional}. Near-field spin--orbit effects have been resolved by photoinduced force microscopy \cite{dong2024nearfield}, and orbital-to-spin angular momentum conversion has been studied with a single silicon nanoparticle \cite{nechayev2019orbital}. Near-field angular momentum also underlies polarization-dependent trapping and rotation of particles near micro- and nanostructures \cite{tsai2014selective,canosvalero2020nanovortex,lu2023polarization,lu2023helicity}. Controlling the sign and strength of the converted angular momentum in such near fields is therefore important for both nano-optics and optical manipulation.

For a structure with continuous rotational symmetry, such as a circular pillar illuminated along its axis, the total angular momentum along the axis is conserved \cite{schwartz2006conservation,haefner2009spin,bliokh2011spin}. The converted OAM is then carried by field components whose phase winds in the same direction as the incident spin, as shown for focusing and scattering systems \cite{bliokh2011spin}, so it has the same sign as the spin. Reversing the near-field OAM therefore seems to require reversing the polarization. Breaking the rotational symmetry lifts this constraint in two ways. First, in chiral structures such as plasmonic spirals, the topological charge of the near-field vortex includes a geometric charge determined by the handedness of the spiral, so the sign of the near-field OAM also depends on the structure \cite{tsai2014selective,kim2010synthesis}. Second, a structure with $N$-fold rotational symmetry allows field components whose angular momentum differs from the incident one by multiples of $N$ \cite{lee2014optical,chen2014symmetry}, some of which carry angular momentum opposite to the spin. In plasmonic plates, such components rotate opposite to the incident field at their resonances \cite{lee2014optical,ichiji2024designing} and can reverse the angular momentum carried away by the scattered light \cite{lee2014optical}, and collective modes of optically bound nanoparticle arrays can reverse the optical torque \cite{han2018crossover,parker2020optical}. These studies focus on individual components, the scattered light, and the torque on particles. Whether symmetry breaking can reverse the net OAM of the near field in an achiral structure has, to our knowledge, remained open.

Here we show that the OAM of the near field can be opposite to the incident spin in achiral structures. A circular pillar serves as the reference. Its near-field OAM follows the incident spin and is enhanced at its resonances, but it never reverses. Breaking the continuous rotational symmetry makes such a reversal possible, and we find it, for example, in triangular, square, and pentagonal silicon pillars and in clusters of circular pillars, in distinct spectral bands. The reversal does not invert the phase winding of the longitudinal field. It arises when higher-order vortices with topological charge opposite to the spin dominate over those with charge of the same sign, and an exact decomposition of the angular momentum into angular harmonics gives a criterion for the reversal. Projections of the simulated fields show that these higher-order components can be carried by modes of the circular pillar with opposite angular momentum. Circularly polarized light cannot excite these modes as long as the continuous rotational symmetry is preserved, but can excite them once it is broken.

\begin{figure*}[t]
\centering
\includegraphics[width=\textwidth]{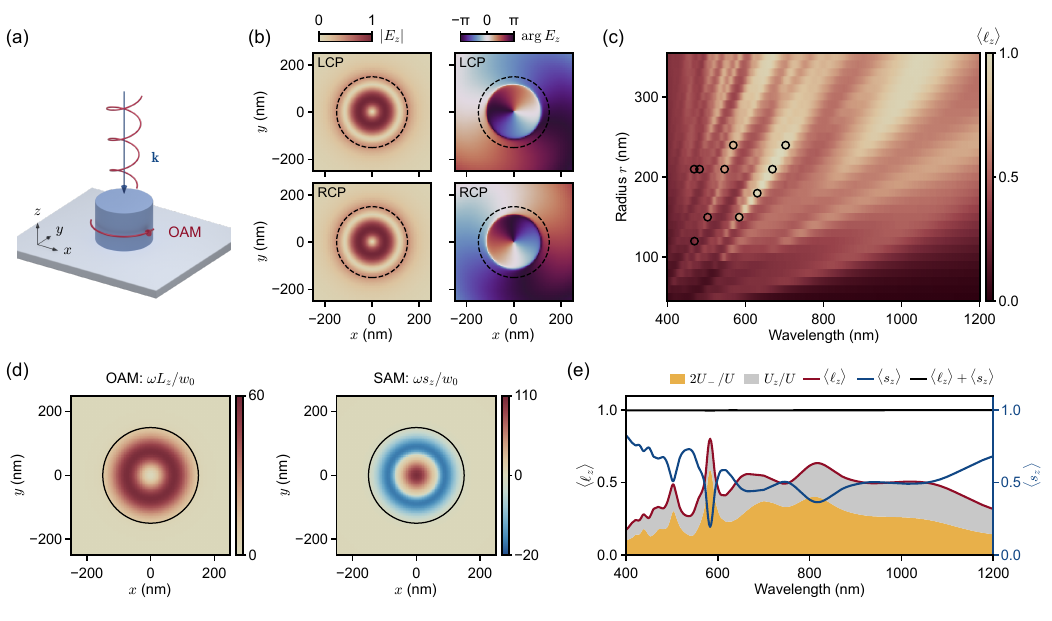}
\caption{\label{fig:circular}Near-field spin-to-orbital angular momentum conversion around a circular silicon pillar.
(a) Schematic. A silicon pillar on a silica substrate in water, illuminated by circularly polarized light along $-z$.
(b) Longitudinal-field amplitude (normalized to its maximum) and phase under LCP (top) and RCP (bottom) illumination.
(c) Normalized OAM $\ellz$ versus wavelength and pillar radius (color scale clipped at 1). Open circles mark the resonance wavelengths of selected $J_z=1$ quasinormal modes.
(d) Normalized orbital and spin angular momentum densities, $\omega L_z/w_0$ and $\omega s_z/w_0$, under LCP illumination, where $w_0=\varepsilon_\mathrm{b}|E_0|^2/2$ is the incident energy density.
(e) Normalized OAM $\ellz$, normalized spin $\sz$, and their sum for $r=\SI{150}{\nano\meter}$. The stacked areas show $2U_-/U$ and $U_z/U$, with $U_-$ and $U_z$ the spin-flipped and longitudinal component energies.
All pillars have height \SI{220}{\nano\meter}. In (b) and (d), $r=\SI{150}{\nano\meter}$, $\lambda=\SI{583.3}{\nano\meter}$, and $z=\SI{50}{\nano\meter}$, with the substrate surface at $z=0$. Circles outline the pillar.
In (c) and (e), the illumination is LCP.}
\end{figure*}

\section{Conversion near a circular pillar}\label{sec:model}\label{sec:single}\label{sec:mechanism}

We first consider a circular silicon pillar, for which the sign of the converted OAM is fixed by the incident spin [Fig.~\ref{fig:circular}(a)]. The pillar, of height $h=\SI{220}{\nano\meter}$ and radius $r=\SI{150}{\nano\meter}$ unless stated otherwise, stands on a silica substrate in water (refractive index $n_\mathrm{b}=1.33$) and is illuminated along $-z$ by a circularly polarized plane wave. The substrate surface is at $z=0$, so the pillar occupies $0<z<\SI{220}{\nano\meter}$. We use the time dependence $e^{-i\omega t}$ and denote left circularly polarized (LCP) light by $\sigma=+1$, with spin along $+z$, and right circularly polarized (RCP) light by $\sigma=-1$. Any other incident polarization is a superposition of these two states (Sec.~S1). We calculate the fields with the finite-difference time-domain (FDTD) method, and the numerical methods are given in the Supplemental Material~\cite{supplemental}.

The longitudinal field reveals the conversion [Fig.~\ref{fig:circular}(b)]. The incident wave has no longitudinal component, but the pillar generates an $E_z$ whose phase winds once around the axis, in opposite directions for LCP and RCP, while its amplitude is the same for both. This phase winding carries OAM, whose sign follows the incident spin.

To quantify the conversion, we use the dual-symmetric canonical angular momentum densities of the near field \cite{Bliokh_2013,bliokh2014extraordinary}. Multiplied by $\omega$, the densities of OAM and spin along the pillar axis are written as
\begin{equation}
\begin{aligned}
\omega L_z&=\tfrac14\operatorname{Im}\!\left[\varepsilon\mathbf{E}^*\!\cdot\partial_\varphi\mathbf{E}+\mu_0\mathbf{H}^*\!\cdot\partial_\varphi\mathbf{H}\right],\\
\omega s_z&=\tfrac14\operatorname{Im}\!\left[\varepsilon(\mathbf{E}^*\times\mathbf{E})_z+\mu_0(\mathbf{H}^*\times\mathbf{H})_z\right],
\end{aligned}
\label{eq:densities}
\end{equation}
where $\varphi$ is the azimuthal angle about the pillar axis, $\partial_\varphi$ acts on the Cartesian components, and $\varepsilon$ is the permittivity at each position, $\varepsilon_\mathrm{b}=\varepsilon_0n_\mathrm{b}^2$ in water and $\varepsilon_0\operatorname{Re}\varepsilon_\mathrm{Si}$ in silicon, with $\varepsilon_\mathrm{Si}$ its complex relative permittivity. We ignore the dispersion correction for silicon \cite{bliokh2017dispersive,picardi2018angular} for simplicity. The anomalous conversion spectral bands reported below persist even with the dispersion correction (Sec.~S2).

We integrate these densities over a near-field region around the pillar and normalize them by the energy in the same region,
\begin{equation}
\ellz=\frac{1}{U}\int_V\omega L_z\,dV,\qquad \sz=\frac{1}{U}\int_V\omega s_z\,dV,
\label{eq:normalized}
\end{equation}
where $w=\tfrac14(\varepsilon|\mathbf{E}|^2+\mu_0|\mathbf{H}|^2)$ is the energy density, $U=\int_Vw\,dV$, and $V$ is a slab within the pillar height that contains the pillar and its near field (Sec.~S2). We call $\ellz$ the normalized OAM. Both quantities are energy-weighted spatial averages of the local ratios $\omega L_z/w$ and $\omega s_z/w$, and in homogeneous water they give the angular momentum per photon in units of $\hbar$. A paraxial vortex beam of topological charge $\ell$ and spin $\sigma$ gives $\ellz\approx\ell$ and $\sz\approx\sigma$.

The rotational symmetry determines which field components carry this angular momentum. In the circular basis $\hat{\mathbf{e}}_\pm=(\hat{\mathbf{x}}\pm i\hat{\mathbf{y}})/\sqrt{2}$, the field $\mathbf{E}=E_+\hat{\mathbf{e}}_++E_-\hat{\mathbf{e}}_-+E_z\hat{\mathbf{z}}$ has three components, which carry spin $s=+1$, $-1$, and 0, respectively. We label them relative to the incident spin. The co-rotating component $E_\sigma$ rotates in the same direction as the incident field ($E_+$ for LCP and $E_-$ for RCP), the spin-flipped component $E_{-\sigma}$ rotates in the opposite direction ($E_-$ for LCP and $E_+$ for RCP), and $E_z$ is longitudinal. Because the pillar and the substrate are rotationally symmetric about $z$, scattering conserves the total angular momentum $J_z$, so the near field has the same $J_z=\sigma$ as the incident wave. A component with spin $s$ has the phase winding $J_z-s$ (Sec.~S1), so each component has a single phase winding. The component $E_\sigma$ has winding 0 and does not depend on $\varphi$, $E_z\propto e^{i\sigma\varphi}$, and $E_{-\sigma}\propto e^{2i\sigma\varphi}$ [Fig.~S4]. The components $H_\sigma$, $H_z$, and $H_{-\sigma}$ of the magnetic field have the same windings \cite{bliokh2011spin}. Substituting these forms into Eq.~\eqref{eq:densities} gives, at every point (Sec.~S1),
\begin{equation}
\omega L_z=\sigma\left(w_z+2w_{-\sigma}\right),\qquad \omega s_z=\sigma\left(w_\sigma-w_{-\sigma}\right),
\label{eq:pointwise}
\end{equation}
where $w_j=\tfrac14(\varepsilon|E_j|^2+\mu_0|H_j|^2)$ is the energy density of component $j=\sigma,-\sigma,z$. The ratio $\omega(L_z+s_z)/w=\sigma$ is fixed by the incident light, so scattering only redistributes angular momentum between its spin and orbital parts. Integrating Eq.~\eqref{eq:pointwise} gives
\begin{equation}
\ellz=\sigma\,\frac{U_z+2U_{-\sigma}}{U},\qquad \ellz+\sz=\sigma,
\label{eq:identity}
\end{equation}
with $U_j=\int_Vw_j\,dV$. Near a circular pillar, the OAM is thus carried by the longitudinal component, $U_z$, with one unit per photon, and by the spin-flipped component, $U_{-\sigma}$, with two units per photon. Because these energies cannot be negative, Eq.~\eqref{eq:identity} restricts $\sigma\ellz$ to the range from 0, with all of the energy in the co-rotating component, to 2, with all of it in the spin-flipped component. A resonance can increase the magnitude of the OAM, but the OAM cannot have the sign opposite to the incident spin. The same holds for ring pillars, which also have continuous rotational symmetry (Sec.~S4).

For $\sigma\ellz=1$, Eq.~\eqref{eq:identity} gives $\sz=0$, so the near field in $V$ carries, on average, one unit of OAM per photon and no net spin, although the local spin density is positive near the axis and negative farther from it [Figs.~\ref{fig:circular}(d) and S6(f)]. Because $\ellz$ is normalized by the near-field energy, this value is not the fraction of the incident optical power that is converted. Because $U=U_\sigma+U_{-\sigma}+U_z$, Eq.~\eqref{eq:identity} also gives $\sigma\ellz-1=(U_{-\sigma}-U_\sigma)/U$, so $\sigma\ellz>1$ when the spin-flipped component carries more energy than the co-rotating component. A similar decomposition into field components is known for focused and scattered beams \cite{bliokh2011spin}. Equation~\eqref{eq:identity} applies it to the electric and magnetic near field of the pillar.

We now apply these relations to the simulated near field. The local densities show how the OAM and the spin are distributed in the pillar [Fig.~\ref{fig:circular}(d)]. Under LCP illumination, the OAM density is positive and forms a ring inside the pillar, whereas the spin density is positive near the axis and negative in a ring around it, where $w_{-\sigma}$ exceeds $w_\sigma$. The corresponding OAM and spin density distributions in the longitudinal cross section (the $xz$ plane) are shown in Fig.~S2.

The normalized OAM spectra obtained from the FDTD simulations show resonances whose wavelengths depend on the pillar radius [Fig.~\ref{fig:circular}(c)]. At $r=\SI{150}{\nano\meter}$, $\ellz$ peaks at 0.80 at a wavelength of \SI{583}{\nano\meter}, and this peak redshifts and strengthens as the radius increases, reaching $\ellz=1.00$ at a wavelength of \SI{628}{\nano\meter} for $r=\SI{180}{\nano\meter}$ and 1.04 at \SI{667}{\nano\meter} for $r=\SI{210}{\nano\meter}$. At $r=\SI{180}{\nano\meter}$ and a wavelength of \SI{628}{\nano\meter}, the spin-flipped and co-rotating energy fractions are nearly equal (both 0.38), so that $\sz\approx0$ and $\ellz\approx1$.

The resonant increase of $\ellz$ comes mainly from the spin-flipped component [Fig.~\ref{fig:circular}(e)]. Over 400--\SI{1200}{\nano\meter}, $\ellz+\sz=1$, in agreement with Eq.~\eqref{eq:identity}. At the \SI{583}{\nano\meter} peak, under LCP illumination ($U_{-\sigma}=U_-$), $\ellz=0.80$ and $\sz=0.19$. The spin-flipped term gives $2U_-/U=0.59$ (74\% of the OAM), and the longitudinal term $U_z/U$ gives the remainder.

These spectra give the response of the pillar under circularly polarized plane-wave illumination. To identify the resonances behind their peaks, we compute the quasinormal modes (QNMs) of the circular pillar on its substrate in water, that is, its eigenmodes with complex frequencies, for $J_z=1$, the only modes that LCP illumination can excite. The $\ellz$ peak at \SI{583}{\nano\meter} lies near a QNM at \SI{583.6}{\nano\meter}. The quality factor of this mode, $Q\approx30$, corresponds to a linewidth of 15--\SI{20}{\nano\meter} (Sec.~S5), comparable to the \SI{20}{\nano\meter} width of the simulated peak (full width at half of its height above the background). The mode is dominated by the spin-flipped component, $U_-/U=0.57$, with $w_-$ in two lobes along the pillar and $w_+$ concentrated near the axis [Fig.~S6(c)]. Its $\ellz$, evaluated from the mode field alone, is 1.40. As the radius increases, the resonance wavelengths of these $J_z=1$ QNMs follow the $\ellz$ peak, as marked by the open circles in Fig.~\ref{fig:circular}(c). To quantify the excitation of this mode, we expand the FDTD-simulated electric field inside the pillar in the ten $J_z=1$ QNMs found between 430 and \SI{600}{\nano\meter}. At \SI{583}{\nano\meter}, the \SI{583.6}{\nano\meter} mode has the largest expansion term, whose integrated intensity is 0.63 of that of the FDTD-simulated field, and the part of the field that the ten modes do not reproduce holds 28\% of the electric-field intensity in the pillar (Sec.~S5). The conversion peak does not coincide with strong scattering, because the scattering efficiency at \SI{583}{\nano\meter} is less than 40\% of its maximum (Sec.~S3).

\begin{figure}[!tbp]
\centering
\includegraphics[width=\columnwidth]{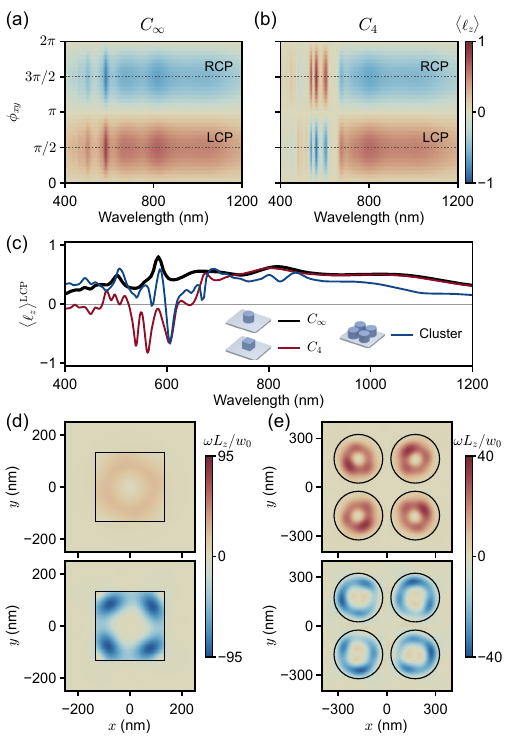}
\caption{\label{fig:overview}Symmetry-enabled anomalous near-field spin-to-orbital angular momentum conversion.
(a),(b) Normalized OAM versus wavelength and relative polarization phase $\phi_{xy}$ for the circular (a) and square (b) pillars.
(c) Normalized OAM spectra under LCP illumination. The cluster curve gives the OAM per pillar in a \SI{350}{\nano\meter} window. The legend includes schematics of a circular pillar ($C_\infty$, $r=\SI{150}{\nano\meter}$), a square pillar ($C_4$, side \SI{266}{\nano\meter}), and a square cluster of four circular pillars (Cluster, each with $r=\SI{150}{\nano\meter}$, edge-to-edge gap $g=\SI{50}{\nano\meter}$).
(d),(e) Normalized OAM density $\omega L_z/w_0$ for the square pillar (d) at $(\lambda,z)=(801.5,65)$~nm (top) and $(562.2,165)$~nm (bottom) and for the cluster (e) at $(582.9,55)$~nm (top) and $(606.6,105)$~nm (bottom). The value of $\ellz$ is positive at the wavelengths of the top maps and negative at those of the bottom maps. Black lines outline the pillars.
All pillars have height \SI{220}{\nano\meter} and stand on silica in water, with the substrate surface at $z=0$.}
\end{figure}

\section{Reversal of the near-field OAM}\label{sec:anomalous}

The results above show that the near-field OAM of the circular pillar is enhanced at its resonances, but $\sigma\ellz\ge0$ at every wavelength and radius. Breaking the continuous rotational symmetry removes this restriction. We compare the circular pillar with an equal-area square pillar (side \SI{266}{\nano\meter}) and with a square cluster of four circular pillars (each of radius \SI{150}{\nano\meter}, edge-to-edge gap $g=\SI{50}{\nano\meter}$) [Fig.~\ref{fig:overview}(c)]. The four pillars of the cluster are equivalent under its fourfold rotational symmetry, so we analyze the angular momentum of one pillar about its own axis (Sec.~S2).

Maps of $\ellz$ versus wavelength and incident polarization show the difference most clearly [Figs.~\ref{fig:overview}(a) and \ref{fig:overview}(b)]. The polarization is set by $\phi_{xy}$, the phase of the $+45^\circ$ incident linear-polarization component relative to the $-45^\circ$ component [Eq.~(S1)], with LCP at $\pi/2$ and RCP at $3\pi/2$. For the circular pillar, $\ellz$ follows the incident spin. It is positive for LCP, negative for RCP, and changes sign only with the polarization. For the square pillar, $\ellz$ has the opposite sign, negative under LCP and positive under RCP, between 525 and \SI{622}{\nano\meter} and, more weakly, below \SI{474}{\nano\meter}. From these maps we find that $\ellz(\phi_{xy})=\ellz^\mathrm{LCP}\sin\phi_{xy}$ for both pillars, which is a consequence of the fourfold rotational and vertical mirror symmetries shared by each pillar and its integration region (Sec.~S1). The LCP spectrum therefore determines each map, and its sign shows whether the OAM follows or opposes the incident spin. We call the conversion anomalous when the near-field OAM is opposite to the incident spin, $\sigma\ellz<0$, that is, $\ellz^\mathrm{LCP}<0$ under LCP illumination. Such a reversal does not violate angular-momentum conservation. The total angular momentum of the light is conserved only if the structure is invariant under rotations about the axis. Otherwise, the structure exchanges angular momentum with the field, so the near field can carry an angular momentum that differs from the incident one.

Under LCP illumination, the square-pillar OAM reaches $-0.82$ near \SI{562}{\nano\meter}, whereas the circular-pillar OAM stays positive at all wavelengths [Fig.~\ref{fig:overview}(c)]. The cluster also reverses the OAM, with a minimum near \SI{606}{\nano\meter}, although each of its pillars is circular, because the neighbors break the rotational symmetry about each pillar axis. At the wavelengths where $\ellz$ is negative, the OAM density becomes negative near the corners of the square pillar and in a ring inside each cluster pillar [Figs.~\ref{fig:overview}(d) and \ref{fig:overview}(e)].

Equation~\eqref{eq:identity} cannot produce these negative values, because $U_z$ and $U_{-\sigma}$ cannot be negative. The reversal therefore requires field components that circularly polarized light cannot excite in the circular pillar, which we identify next.

\section{Angular-momentum channels}\label{sec:results}

\begin{figure}[!tbp]
\centering
\includegraphics[width=\columnwidth]{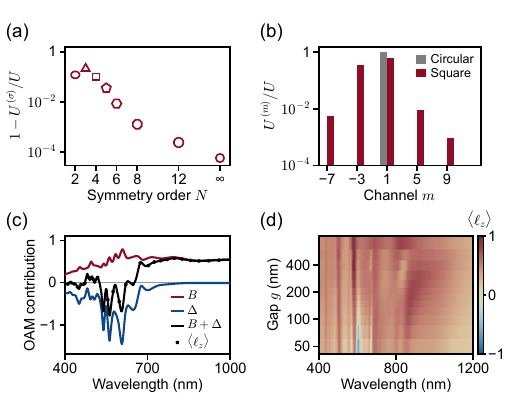}
\caption{\label{fig:symmetry}Angular-momentum channels and their contributions to anomalous conversion under LCP illumination.
(a) Leakage $1-U^{(\sigma)}/U$ from the incident channel $m=\sigma=1$, averaged over the sampled wavelengths from 450 to \SI{1200}{\nano\meter}. Marker shapes represent the cross sections, namely an ellipse for $N=2$, regular polygons for finite $N\ge3$, and a circle for $N=\infty$.
(b) Channel energy fractions $U^{(m)}/U$ for the circular and square pillars at \SI{606.1}{\nano\meter}.
(c) OAM of the square pillar and its two parts, with $B=(U_z+2U_-)/U$ and $\Delta=\sum_m(m-1)U^{(m)}/U$. The black line shows $B+\Delta$, and black dots show $\ellz$ integrated directly over the circular region of radius \SI{350}{\nano\meter} used for the channels.
(d) Per-pillar normalized OAM in a square cluster of four circular pillars versus wavelength and edge-to-edge gap $g$.
Panels (a)--(c) use $\varepsilon_\mathrm{b}$ in the densities (Sec.~\ref{sec:results}), and (d) uses the permittivity of Eq.~\eqref{eq:densities}. All pillars have height \SI{220}{\nano\meter}. Each circular pillar, including each of the four cluster pillars, has radius \SI{150}{\nano\meter}, and the noncircular cross sections have the same area.}
\end{figure}

\label{sec:symmetry}
Without continuous rotational symmetry, the field no longer has a single value of $J_z$. We expand it in angular harmonics with integer index $m$ about the pillar axis. Each harmonic, which we call a channel, carries total angular momentum $m$ per photon, and the incident channel is $m=\sigma=\pm1$. In a noncircular pillar, the permittivity depends on $\varphi$, so the densities of Eq.~\eqref{eq:densities} contain products of different channels. To separate the channels exactly, we replace the permittivity in the densities by the constant $\varepsilon_\mathrm{b}$ everywhere, including inside silicon, when we evaluate the channel quantities defined below ($U^{(m)}$, $B$, and $\Delta$) and the projections of Sec.~\ref{sec:negative}. This replacement changes the values of $\ellz$, but every reversal band persists (Sec.~S2 and Table~S1). Unless stated otherwise, the values of $\ellz$ below use the permittivity of Eq.~\eqref{eq:densities}. Integrating over a circular region centered on the axis (of radius \SI{350}{\nano\meter} for the isolated pillars) then removes the products of different channels and gives (Sec.~S1)
\begin{equation}
\Delta\equiv\ellz+\sz-\sigma=\sum_m(m-\sigma)\,\frac{U^{(m)}}{U},
\label{eq:sumrule}
\end{equation}
where $U^{(m)}$ is the energy in channel $m$ and $\Delta$ is the angular-momentum mismatch relative to the incident channel. For $N$-fold rotational symmetry ($C_N$), only the channels $m=\sigma+kN$, with integer $k$, can be excited, so each channel contributes to $\Delta$ in proportion to its change in angular momentum, $m-\sigma=kN$. The quantity $\Delta$ describes the near field in the chosen region and does not directly give the torque exerted on the structure.

Because $\sz=\sigma(U_\sigma-U_{-\sigma})/U$ holds for any field, Eq.~\eqref{eq:sumrule} can be rewritten as
\begin{equation}
\ellz=\sigma B+\Delta,\qquad B\equiv\frac{U_z+2U_{-\sigma}}{U}\ge0.
\label{eq:balance}
\end{equation}
This decomposition splits the OAM into two parts. The first, $\sigma B$, is the contribution of the longitudinal and spin-flipped energies, as in Eq.~\eqref{eq:identity}, and always has the sign of the spin. The second, $\Delta$, is the shift produced by all channels other than $m=\sigma$. The OAM therefore reverses, $\sigma\ellz<0$, exactly when $\sigma\Delta<-B$. For LCP illumination, $\sigma=1$ and $\Delta=\sum_m(m-1)U^{(m)}/U$, so the channels with $m<1$ make $\Delta$ negative and those with $m>1$ make it positive. We call the channels with $m<1$ opposite channels because they lower the angular momentum relative to the incident channel. The OAM reverses when $\sum_{m<1}(1-m)U^{(m)}/U$ exceeds $B+\sum_{m>1}(m-1)U^{(m)}/U$. Symmetry determines which channels are allowed, whereas their energies depend on the optical response of the structure.

We first examine how much energy a noncircular boundary transfers out of the incident channel. We consider pillars whose cross sections are regular $N$-gons of the same area as the $r=\SI{150}{\nano\meter}$ circular pillar ($C_N$ symmetry, with an ellipse for $N=2$). We quantify the transfer by the leakage $1-U^{(\sigma)}/U$, the fraction of the near-field energy outside the incident channel, which falls steeply with $N$ for $N\ge3$ [Fig.~\ref{fig:symmetry}(a)]. Averaged over 450--\SI{1200}{\nano\meter}, it is 0.21 for $N=3$, 0.10 for $N=4$, 0.036 for $N=5$, 0.0087 for $N=6$, 0.0013 for $N=8$, and $2\times10^{-4}$ for $N=12$. For the circular pillar it vanishes, within a numerical accuracy of about $10^{-4}$. For the ellipse ($N=2$), the leakage is 0.12, but $\ellz$ only approaches zero (minimum $-0.012$ near \SI{581}{\nano\meter}), because its nearest opposite channel, $m=-1$, changes the angular momentum by only $-2$ per photon, so that $\Delta$ barely cancels $B$.

We next ask which allowed channel produces the reversal. For the square pillar, it comes mainly from one opposite channel, $m=-3$ [Figs.~\ref{fig:symmetry}(b) and \ref{fig:symmetry}(c)]. Under LCP illumination, the nearest allowed channels $m=1\pm N$ shift the angular momentum by $\pm N$. At \SI{606}{\nano\meter}, 36\% of the energy in the circular region occupies $m=-3$ and 62\% remains in $m=1$, whereas every other channel holds at most 1\% (note the logarithmic axis of Fig.~\ref{fig:symmetry}(b)). A component with spin $s$ in channel $m$ has the phase winding $m-s$, as in Sec.~\ref{sec:mechanism}. The $E_+$, $E_-$, and $E_z$ components of the channel $m=-3$ therefore have phase windings $-4$, $-2$, and $-3$, so every component contributes negative OAM. The reversal thus does not turn the winding $+1$ of $E_z$ into $-1$. Near 562 and \SI{606}{\nano\meter}, $E_z$ still has the winding $+1$ close to the axis. Farther from the axis, the windings $-3$ in $E_z$, $-4$ in $E_+$, and $-2$ in $E_-$ of the channel $m=-3$ dominate at some radii. The sum of the windings weighted by energy is therefore negative, and the OAM reverses.

At \SI{562}{\nano\meter}, $B=0.60$ and $\Delta=-1.38$, giving $\ellz\simeq-0.78$ with $\varepsilon_\mathrm{b}$ in the circular region. The $m=-3$ channel holds 35\% of the energy and contributes $-1.39$ to $\Delta$, whereas all other channels together contribute $+0.01$. The band from 525 to \SI{622}{\nano\meter} contains further minima of $-0.71$ near \SI{540}{\nano\meter} and $-0.66$ near \SI{605}{\nano\meter}.

The triangular and pentagonal pillars also reverse the OAM (Sec.~S4). The triangular pillar reaches $-0.37$ near \SI{426}{\nano\meter} and $-0.34$ near \SI{509}{\nano\meter}, and the pentagonal pillar reaches $-0.63$ near \SI{492}{\nano\meter}. For $N\ge6$, $\ellz$ stays above 0.10.

Neighbors provide a second way to transfer energy into the opposite channels. A square cluster of four circular pillars has global $C_4$ symmetry, but each pillar has no rotational symmetry about its own axis, so all channels about its axis are allowed. The per-pillar $\ellz$ becomes negative near \SI{606}{\nano\meter} for edge-to-edge gaps $g\le\SI{125}{\nano\meter}$ [Fig.~\ref{fig:symmetry}(d)]. At $g=\SI{50}{\nano\meter}$ it reaches $-0.67$ and is negative over 597--\SI{619}{\nano\meter}. As the gap grows, it recovers toward the isolated-pillar value at \SI{606}{\nano\meter}, about 0.58. A channel decomposition about the pillar axis places 37\% of the energy in $m=-3$ near \SI{606}{\nano\meter} (Sec.~S4). Both a noncircular boundary and neighboring pillars thus transfer energy into the same opposite channel, $m=-3$.

\begin{figure}[!tbp]
\centering
\includegraphics[width=\columnwidth]{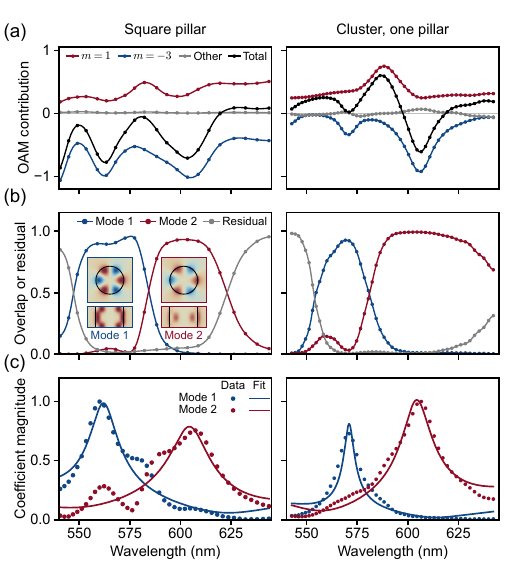}
\caption{\label{fig:mminus3}Modal analysis of anomalous near-field spin-to-orbital angular momentum conversion in the square pillar (left) and one pillar of the $g=\SI{50}{\nano\meter}$ cluster (right) under LCP illumination.
(a) OAM contributions from $m=1$, $m=-3$, and all other channels with $|m|\le12$, normalized by the energy in a cylinder of radius \SI{125}{\nano\meter} about the pillar axis, together with their sum.
(b) Single-mode squared overlaps $\eta_1$ and $\eta_2$ with the $m=-3$ part of the FDTD-simulated field, and the residual $R$ of the projection onto both modes.
(c) Normalized magnitudes of the complex projection coefficients (dots) and their complex single-resonance fits (lines), each with a linear background and an adjustable resonance frequency and linewidth.
In (b) and (c), Mode 1 (blue) and Mode 2 (red) are the reference circular-pillar modes at 570.0 and \SI{604.9}{\nano\meter}, respectively. The insets in (b) show, for each mode, $\operatorname{Re}E_z$ in a horizontal plane (top), at $z=\SI{55}{\nano\meter}$ inside the pillar for Mode 1 and at $z=\SI{225}{\nano\meter}$, just above its top face, for Mode 2, and $|\mathbf{E}|^2$ in a vertical plane through the axis (bottom), each normalized to its maximum. Black lines outline the circular pillar. The densities and overlaps use $\varepsilon_\mathrm{b}$ (Sec.~\ref{sec:results}).}
\end{figure}

\section{Resonant origin of the opposite channel}\label{sec:negative}

The opposite channel is strong only in certain spectral bands, which suggests a resonant origin. The circular pillar itself supports modes with $J_z=-3$. Between 550 and \SI{640}{\nano\meter} we find two, Mode 1 at \SI{570.0}{\nano\meter} and Mode 2 at \SI{604.9}{\nano\meter}, with quality factors of 48 and 30. Under circularly polarized illumination along the axis, the circular pillar cannot excite them, because $J_z=\sigma$ is conserved. The corners of the square pillar and the neighbors in the cluster couple the incident channel to $m=-3$ and thus make these modes accessible.

The reversal in both structures comes mainly from the $m=-3$ channel [Fig.~\ref{fig:mminus3}(a)]. For this comparison and the projections below, we integrate over a cylinder of radius \SI{125}{\nano\meter} about the pillar axis, which lies inside the silicon in both structures, and normalize each channel's OAM by the total energy in this cylinder. In both structures the negative $m=-3$ contribution exceeds the positive $m=1$ contribution in magnitude in the bands where the total in Fig.~\ref{fig:mminus3}(a) is negative, while the other channels contribute little.

To test whether the two modes describe the $m=-3$ field, we project the $m=-3$ channel of the FDTD-simulated field of the square pillar and the cluster onto Modes 1 and 2. We denote the electric and magnetic fields of this channel together by $d$, and those of mode $j$ ($j=1,2$) by $p_j$. With the dual-symmetric overlap integral $\langle p,d\rangle=\int(\varepsilon_\mathrm{b}\mathbf{E}_p^*\!\cdot\mathbf{E}_d+\mu_0\mathbf{H}_p^*\!\cdot\mathbf{H}_d)\,dV$ over this cylinder and $\|d\|^2=\langle d,d\rangle$, we evaluate
\begin{equation}
\eta_j=\frac{|\langle p_j,d\rangle|^2}{\|p_j\|^2\|d\|^2},\qquad
R=\frac{\|d-\sum_jc_jp_j\|^2}{\|d\|^2},
\label{eq:negative_projection}
\end{equation}
where $\eta_j$ is the squared overlap with mode $j$ alone and $R$ is the fraction of $\|d\|^2$ that the two modes together do not reproduce, with the complex coefficients $c_j$ chosen to minimize it. These overlaps measure the similarity of field shapes within the $m=-3$ channel and are not fractions of the energy of the full field.

The two modes describe the opposite channel near 562--570 and \SI{606}{\nano\meter} [Fig.~\ref{fig:mminus3}(b)]. For the square pillar, the overlap with Mode 1 is 0.90 at \SI{562.2}{\nano\meter}, and that with Mode 2 is 0.93 at \SI{606.1}{\nano\meter}, with two-mode residuals of 0.025 and 0.058. For the cluster, the Mode 1 overlap is 0.93 at \SI{569.1}{\nano\meter} and the Mode 2 overlap is 0.99 at \SI{605.5}{\nano\meter}, with residuals below 0.01. The minimum of the square pillar near \SI{540}{\nano\meter} lies outside this description ($R=0.85$ at \SI{540.5}{\nano\meter}), so its $m=-3$ field involves $J_z=-3$ modes outside the investigated wavelength range (550--\SI{640}{\nano\meter}) or a nonresonant response.

The projection coefficients also show resonant behavior [Fig.~\ref{fig:mminus3}(c)]. We fit each complex coefficient with a single resonant term and a linear complex background, allowing the resonance frequency and linewidth to adjust while keeping the spatial modes of the circular pillar. The fits reproduce the coefficients with normalized squared errors below 0.18 (Table~S3). Holding the resonance frequency and linewidth at their circular-pillar values raises the Mode 1 error by a factor of 4.5 for the square pillar and 2.4 for the cluster, whereas Mode 2 of the square pillar changes little. The spatial overlaps and spectral fits therefore show that the opposite channel is resonantly enhanced by modes of the circular pillar. The corners or the neighbors perturb these modes and shift their complex resonance frequencies, mainly in wavelength for Mode 1 of the square pillar and in linewidth for Mode 1 of the cluster (Sec.~S5). The phases of the coefficients and of their fits are shown in Fig.~S7.

The same analysis applies to other cross sections, where the opposite channel depends on the symmetry. The nearest opposite channel is $m=1-N$, and it dominates the reversal near \SI{509}{\nano\meter} in the triangular pillar and near \SI{492}{\nano\meter} in the pentagonal pillar. In the triangular pillar, $m=-2$ holds 31\% of the energy in the \SI{350}{\nano\meter} circular region and contributes $-0.94$ to $\Delta$, and in the pentagonal pillar, $m=-4$ holds 20\% and contributes $-1.02$. In the pentagonal pillar, the channel $m=-4$ is carried by a mode of the circular pillar, as for the square pillar. Its field at \SI{492}{\nano\meter} has a squared overlap of 0.94 with a $J_z=-4$ mode at \SI{493.4}{\nano\meter} ($Q=29$). In the triangular pillar, none of the six $J_z=-2$ modes that we computed dominates the $m=-2$ field at \SI{509}{\nano\meter} (largest overlap 0.41, Sec.~S5), so these modes do not identify the origin of the opposite channel. For these equal-area pillars with $N\ge6$, the opposite channel holds less than 5\% of the energy, and the OAM does not reverse.

\section{Discussion}\label{sec:discussion}

The decomposition of Eq.~\eqref{eq:balance} separates the roles of symmetry and resonance. Continuous rotational symmetry keeps the field in the incident channel and therefore excludes OAM opposite to the spin. Discrete symmetry opens additional channels, $m=\sigma+kN$, as known for scattering by plasmonic particles \cite{lee2014optical} and for nonlinear harmonics \cite{chen2014symmetry}, and resonances determine how much energy they receive \cite{lee2014optical,ichiji2024designing}. Equation~\eqref{eq:balance} turns this separation into an exact criterion for the reversal of the near-field OAM, $\sigma\Delta<-B$, in which symmetry decides which channels enter $\Delta$ and resonances decide their energies. In the square and pentagonal pillars and the cluster, the reversal occurs where resonances of the opposite channel ($m=-3$ or $-4$) make its negative OAM contribution exceed the positive contributions of the other channels. The criterion itself, however, does not require a resonance.

These conclusions depend on how the OAM is evaluated, because the normalized OAM changes with the integration region and with the permittivity in the densities. The circular-pillar relations hold at every point, and therefore in any region and for any of these permittivities. The exact channel decomposition instead requires a region and a permittivity that are invariant under rotations about the chosen axis, which is why Sec.~\ref{sec:results} uses the constant $\varepsilon_\mathrm{b}$. The reversal bands persist with $\varepsilon_\mathrm{b}$ everywhere and with the dispersive permittivity of silicon (Sec.~S2 and Table~S1). Because the substrate preserves these rotations and the vertical mirror planes, all symmetry statements hold with the substrate. When the integration is restricted to the surrounding water, the OAM still reverses for the square pillar and the cluster, whereas the reversal of the triangular and pentagonal pillars comes mainly from the field inside silicon (Table~S1).

For the square pillar and the cluster, anomalous conversion can be tested by measuring the polarization-resolved complex near field and decomposing it into channels. Photoinduced force microscopy has already resolved near-field spin--orbit effects~\cite{dong2024nearfield} and could probe the component energies that enter $B$ in Eq.~\eqref{eq:balance}, whereas $\Delta$ also requires the phase of the near field. The mechanical effect of the near-field OAM on a nanoparticle also depends on the particle response and on its coupling to the structure, and lies beyond the scope of this work.

\section{Conclusion}\label{sec:conclusion}

We have shown that breaking the continuous rotational symmetry of a nanostructure allows the near-field OAM to be opposite to the incident spin. Around a circular silicon pillar, the converted OAM always follows the spin, and resonances with strong spin-flipped fields raise the normalized OAM to about 1. Triangular, square, and pentagonal pillars and clusters of circular pillars instead reverse the OAM in distinct spectral bands. An exact channel decomposition shows that the reversal requires the opposite channels to dominate over the components whose angular momentum has the sign of the spin. These channels can be carried by modes of the circular pillar that the broken symmetry makes accessible.

Symmetry thus determines which channels the near field can contain, while resonances set their strengths. The same decomposition applies to other dielectric structures under illumination with definite angular momentum. Choosing the symmetry and the resonances of a nanostructure could therefore control the sign and the strength of optical angular momentum at subwavelength scales.

\begin{acknowledgments}
This work was supported by the National Natural Science Foundation of China (Grant No. 12674397) and the Fundamental Research Funds for the Central Universities (Grant No. 2042026fg0053).
\end{acknowledgments}

\section*{Data availability}
The data that support the findings of this article are not publicly available. The data are available from the authors upon reasonable request.
\par

\nocite{Evlyukhin:13,alaee2018electromagnetic,lalanne2018light,palik1985handbook,zambranapuyalto2013dual}

\bibliographystyle{apsrev4-2}
\bibliography{refs}
\end{document}